\documentclass[conference]{IEEEtran}
\IEEEoverridecommandlockouts
\usepackage{cite}
\usepackage{amsmath,amssymb,amsfonts,amsthm}
\usepackage{algorithm}
\usepackage{algpseudocode}
\makeatletter
\def\ALG@special@indent{%
    \ifdim\ALG@thistlm=0pt\relax
        \hskip-\leftmargin
    \else
        \hskip\ALG@thistlm
    \fi
}
\newcommand{\Input}[1]{\item[]\noindent\ALG@special@indent \textbf{Input:}\ #1}
\newcommand{\Output}[1]{\item[]\noindent\ALG@special@indent \textbf{Output:}\ #1}

\usepackage{mathtools}

\usepackage{url}
\usepackage{hyperref}
\hypersetup{
    colorlinks=true,
    linkcolor=blue,
    filecolor=magenta,
    urlcolor=blue,
    }

\usepackage{graphicx}
\usepackage{textcomp}
\usepackage{xcolor}
\usepackage{comment}
\def\BibTeX{{\rm B\kern-.05em{\sc i\kern-.025em b}\kern-.08em
    T\kern-.1667em\lower.7ex\hbox{E}\kern-.125emX}}

\xdefinecolor{DarkGreen}{cmyk}{0.8,0,0.8,0.3}

\newcommand{\p}{\mathbf{p}}
\newcommand{\D}{\mathbf{D}}

\newtheorem{definition}{Definition}
\newtheorem{remark}{Remark}
\newtheorem{example}{Example}
\newtheorem{prop}{Proposition}
\begin{document}

\title{Fault Independence in Blockchain \thanks{Preprint to appear in
    the
    \href{https://dsn2023.dei.uc.pt/calls_cfp-disrupt.html}{Disrupt
      Track} of DSN’23.}
  \thanks{This work was supported by the
    Australian Research Council (ARC) under project
    \href{https://dataportal.arc.gov.au/NCGP/Web/Grant/Grant/DE210100019}{DE210100019}.}}
\author{\IEEEauthorblockN{Jiangshan Yu} \IEEEauthorblockA{
    \textit{Monash University}} }

\maketitle

\begin{abstract}
  Byzantine Fault-Tolerant (BFT) protocols have been proposed to
  tolerate malicious behaviors in state machine replications. With
  classic BFT protocols, the total number of replicas is known and
  fixed a priori. The resilience of BFT protocols, i.e., the number of
  tolerated Byzantine replicas (denoted $f$), is derived from the
  total number of replicas according to the quorum theory.

  To guarantee that an attacker cannot control more than $f$ replicas,
  so to guarantee safety, it is vital to ensure fault independence
  among all replicas. This in practice is achieved by enforcing
  diverse configurations of replicas, i.e., each replica has a unique
  configuration, avoiding $f$ fault compromises more than $f$
  replicas.

  While managing replica diversity in BFT protocols has been studied
  in permissioned environments with a small number of replicas, no
  prior work has discussed the fault independence in a permissionless
  environment (such as public blockchains) where anyone can join and
  leave the system at any time. This is particularly challenging due
  to the following two facts. First, with permissionless environment,
  any one can join as a replica at any time and no global coordinator
  can be relied on to manage replica diversity. Second, while great
  progress has been made to scale consensus algorithms to thousands of
  replicas, the replica diversity cannot provide fault independence at
  this scale, limiting practical and meaningful resilience.

  This paper provides the first discussion on the impact of fault
  independence on permissionless blockchains, provides discussions on
  replica configuration diversity, quantifies replica diversity by
  using entropy, and defines optimal fault independence.
\end{abstract}


\section{Introduction}
State Machine Replication (SMR) is a classical technique for
implementing reliability and resilience services. Many critical
infrastructure and Internet services, such as Google Spanner,
implement Crash Fault-Tolerant State Machine Replication (CFT-SMR) to
ensure that their services can continue in the presence of crash
failures, in which machines stop working due to faults such as
hardware failure, software failure, network failure or power
failure. With increasing concerns of cyber-attacks, Byzantine
Fault-Tolerant (BFT) protocols are considered to tolerate arbitrary
faults (such as corruption and intrusions), where machines running BFT
protocols can behave arbitrarily. For example, Boeing 777 and Boeing
787 implement BFT protocols in their Aircraft Information Management
System~\cite{Yeh2001SafetyCA}.

All these CFT and BFT protocols are implemented and executed in
well-controlled SMR environments, where the faults of different
machines are assumed to be independent through diverse replicas. For
clarity, we consider replicas as machines running fault tolerant
protocols to replicate the same state machine. The replica diversity
in a permissioned environment can be implemented by enforcing machines
using different Commercial-Off-The-Shelf components, such as
operating system, external database, and critical crypto and fault
tolerant libraries~\cite{GarciaB019}. In this way, the probability of
the same fault affecting multiple machines is reduced.

Permissionless blockchains~\cite{Vukolic15} are one of the most
popular applications deploying BFT protocols. They introduce a
different system model where any participant can join and leave at any
time. Such an environment makes diversity management very challenging,
due to the lack of trust in individual participants and the lack of a
global diversity manager. While the following discussion is generic
and applies to different permissionless blockchains, for the ease of
presentation we use Bitcoin~\cite{Nakamoto08} as an example in this
paper.

Bitcoin~\cite{Nakamoto08} introduces a different fault model. Rather
than considering the number of tolerated faulty machines, it argues
that the system is secure if an attacker can only control a minority
of hash power. Machines with hash power to perform the Nakamoto
consensus are also called miners~\cite{Nakamoto08}. Following
conventional naming, in the rest of this paper, we use the term
``miner'' to specify replicas in the permissionless blockchains.

Existing works have studied the upper bound of Byzantine hash power the
system can tolerate under different system models~\cite{EyalSirer14,
  NegyRS20, CaoDY23}. In addition, the practicality of such an
assumption (e.g., honest majority) held in practice has also been
analysed and challenged~\cite{Bonneau16, YuKDV19, HanSYL021}.

All the prior works only consider the theoretical bounds of tolerated
Byzantine hash power and factors to break the assumption by gaining
more hash power. For example, prior works consider possibilities of
renting hash power from cloud services or maintaining a mining pool of
distributed machines~\cite{Bonneau16,HanSYL021}. However, to the best
of our knowledge, no work has considered the possibility of a single
fault affecting multiple machines, leading to an attacker controlling
a large amount of honest miners in permissionless environments. This
may happen, for example, if the same operating system or application
software (such as mining software or blockchain wallet) used by
multiple machines has an exploitable vulnerability (such as zero-day
vulnerability). Other examples include exploitable vulnerabilities in
trusted hardware components (such as Intel SGX), that are required by
hybrid fault-tolerant protocols for
blockchain~\cite{Intel-PoET,DecouchantKRY22}. Even though
vulnerabilities can be patched, there exists a vulnerability window
due to the latency in patching
vulnerabilities~\cite{YuR15,BitcoinCore-CVE}.

\textbf{Contribution.} We provide a first step in exploring the impact
of fault independent on permissionless blockchains with the following
two contributions. First, we provide a discussion on replica diversity
and configuration discovery. Second, we propose to use entropy to
quantify replica diversity and define optimal fault independence,
which leads to three propositions on fault independence and
resilience. We hope this would serve as a call for the dependability
and fault-tolerant community to study this critical yet overlooked
challenge.

\section{System and Fault Models}

\subsection{System model}
\label{sec:sysmodel}

We consider a permissionless blockchain environment
where any participant can join and leave at any
time. Let $n_t$ be the total \textit{voting power} in the system at
time $t$. We define \textit{voting power} as an abstraction
representing the total amount of valid voting power units. For
example, for BFT protocols with a fixed number of replicas, $n_t$
represents the total number of replicas at time $t$. For Bitcoin,
$n_t$ represents the total computational power (measured by hashrate)
on the Bitcoin network at time $t$. For permissionless protocols with
membership selection to form a consensus
committee~\cite{abs-1908-08316}, the voting power represents the total
voting power of the committee, as all participants outside of the
consensus committee do not have valid voting power.

A participant with voting power is called a ``\textit{replica}''.
Each replica consists of a machine running a stack of software, where
system software (i.e., operating systems) manages machine hardware and
supports application software (such as implementations of
blockchains).

\subsection{Adversary model}
\label{sec:advmodel}

We consider Byzantine faults, which enables an attacker to arbitrarily
delay, drop, re-order, insert, or modify messages. The total number of
faults is measured and quantified by the amount of affected voting
power. The upper bound $f$ of tolerated faults 
is protocol specific.

We assume the security of the used cryptographic primitives and
protocols, but \textbf{not} their implementations. For example, an
attacker may compromise a replica if the deployed crypto library of
the replica is flawed, but the attacker cannot compromise other
replicas in the same way, if their deployed version of the crypto
library has no exploitable vulnerabilities.

We consider a diverse vulnerabilities leading to Byzantine faults. Let
$k_t$ be the total number of diverse vulnerabilities at time $t$, and
$f^i_t$ be voting power (out of $n_t$) affected by the Byzantine fault
due to the $i$-th vulnerability at time $t$.

\begin{remark}
  Our work can be easily extended to a hybrid model considering
  multiple types of faults (e.g., a mix of Byzantine faults and crash
  faults), by using different variables/parameters to represent the
  total number of different types of faults. In addition, while faults
  can be detected and patched, they do not have any impact on this
  work as the attacks happen during the vulnerability windows.
\end{remark}

\subsection{The challenge}
\label{sec:goal}
To guarantee system security, it is essential to ensure that the total
number of Byzantine faults does not exceed the resilience ($f$
replicas) of the system, i.e.,
$\forall t,\ f\geq \sum_{i=1}^{k_t}f^i_t$.



However, this is extremely challenging to guarantee in permissionless
blockchains. Therefore, the open challenge is identifying efficient
ways to enforce the above equation in a permissionless environment, at
least with a high probability. This introduces the following
sub-challenges, which we provide discussions on the possible solutions
in the next sections:

\begin{itemize}
\item \emph{Challenge 1}: Replica configuration discovery in
  permissionless environments.

\item \emph{Challenge 2:} Fault independence quantification. Assuming
  that the configuration distribution of replicas is known, how to
  measure replica diversity to quantify fault independence is
  challenging.
\end{itemize}

\section{Replica diversity and configuration discovery}
\label{sec:diversity}

This section provides a brief discussion on the replica diversity and configuration discovery.

\subsection{Replica diversity}
\label{sec:Replica_diversity}

We consider three main components of a replica, including trusted
hardware\footnote{Other non-trusted hardware may also fail, but they
  are not an anchor of trust and not as significant when considering
  Byzantine faults.}, system software, and application software.

\textbf{Trusted hardware.} Many blockchain systems rely on trusted
hardware for executing consensus or protecting privacy of smart
contracts. For example, Intel's Hyperledger Sawtooth requires Software
Guard Extension (SGX) to perform PoET as a time-lottery-based
consensus algorithm, allowing replica in the network to vote with
equal chance~\cite{Intel-PoET, sawtooth}. Another example is
Damysus~\cite{DecouchantKRY22}, where trusted hardware components are
employed to improve the performance of streamlined consensus
algorithms. However, trusted hardware are also vulnerable to attacks
(e.g., a recent survey on SGX attacks~\cite{SGX-attacks}). Having
diversity of trusted hardware would help to improve failure
independence.

\begin{remark}
The diversity of trusted hardware is limited, given
the little choices of hardware-assisted isolated execution
environments. One possibility is to implement software protocols to
provide post device compromise security~\cite{YuR15, Yu16,
  MilnerCYR17, YuRC18}. Further discussions on how to detect the
compromise of endpoint trusted hardware are vital for dependable
blockchains but orthogonal to this work.
\end{remark}

\textbf{System software.} Operating system is arguably the heaviest
component, in terms of complexity and lines of code, and the most
targeted component. Given the importance of the operating system,
there are many alternative sub-components to provide diversity in
operating system. Lazarus~\cite{GarciaB019} is a tool to automatically
manage the diversity of operating systems for BFT protocols. While
Lazarus cannot be directly used to manage diversity of operating
systems in permissionless blockchains, we refer readers to their work
for more detailed discussion on the diversity of operating systems.

\textbf{Application software.} Using different
Commercial-Off-The-Shelf (COTS) application software in the software
stack~\cite{DeswarteKL98} can improve fault independence. Out of many
COTS components, such as databases and web browsers, the application
software components that most directly related to blockchain
dependability are the blockchain software. Two of the most important
modules of blockchain software, in terms of security and fault
tolerance, are the key/account management module and consensus module.

Wallet is commonly used to manage keys and accounts. Loosely speaking,
a wallet can be a build-in wallet that is included in the software of
a full node, or it can be a third party software managing private keys
for users. When using a build-in wallet, there is a limited number of
choices (e.g., implementing the same module with different languages
or dependencies), leading to limited (or even no) variety. Third party
wallet has many types, including mobile wallets, web wallets, desktop
wallets, and hardware wallets. However, they are normally designed for
end users rather than for replicas.

Another common way for end users to manage keys is the use of a
delegation (such as exchange platforms), which often manages the
private keys on behave of users. The users only have a pair of user
name and password of the third party service, rather than the actual
public key and private key associated to the account, to claim the
ownership. This allows exchange platforms to gain access to a large
volume of stake, becoming an oligopoly, leading to safety concerns in
the (potentially delegated) proof-of-stake based blockchains. This
also reduces the diversity of replicas, as a potentially large number
of replicas are represented by a single delegate.

Improving the diversity of consensus module, such as N-version BFT
library~\cite{Avizienis85}, is a known open challenge due to its
cost~\cite{GarciaB019}. Possible ways to circumvent this challenge,
protocols with proactive security~\cite{CastroL02, SousaBCNV10,
  DistlerPSRK11, YuRC17, VassantlalAFB22} and
self-stabilization~\cite{Dijkstra74} can be considered to reduce the
potential risk.

Similar to key management, third party software for the consensus
module are also available for users to ``delegate'' the voting
power. For example, mining pool operators in Bitcoin attract and
manage the mining power of distributed participants, leading to an
oligopoly\footnote{The top 10 mining pools in Bitcoin in total possess
  over 96\% mining power (7 day average), where the largest mining
  pool, i.e., Foundry USA, controls over 34\% mining power.
  https://www.blockchain.com/explorer/charts/pools (as of 02 February
  2023.)}. Possible solutions include the design of Non-outsourceable
mining algorithms~\cite{MillerKKS15,HanLY20} and decentralized mining
pools~\cite{LuuVTS17}.

\subsection{Configuration discovery}
\label{sec:configuration_discovery}
We consider the use of remote attestation to discover the
configuration of a replica. The three main components of a replica,
namely its trusted hardware, the system software, and the application
software, can be attested by using remote attestation through trusted
computing. For example, Trusted Platform Modules (TPMs) and Trusted
Execution Environments (TEEs) support remote attestation of the
replica system software and application software. As of Jan 2023, The
\emph{Trusted Computing Group} has certified 41 TPM
products\footnote{https://trustedcomputinggroup.org/membership/certification/tpm-certified-products/},
including 5 products on the latest specification (TPM 2.0, Revision
01.59). Similarly, TEE implementations are also supported by several
hardware technologies, such as \emph{ARM TrustZome}, \emph{Intel
  Software Guard Extensions (SGX)}, \emph{IBM Secure Service
  Container}, and \emph{AMD Platform Security Processor
  (PSP)}. Services, such as Microsoft Azure Attestation, are also
available to provide a unified attestation solution.

\begin{remark}
  \textbf{Additional concerns.} When using remote attestation, it is
  essential to associate the secret key for attestation and the secret
  key for authenticating a vote, proving that a vote indeed comes from
  a replica with the attested configuration. This can be done in
  different ways, for example, by using the TEE to create a vote as in
  Damysus BFT~\cite{DecouchantKRY22}. In addition, the privacy of
  replica configuration should also be protected, as otherwise it
  provides attackers a clear target when new vulnerabilities are
  exposed.
\end{remark}

\section{Measuring replica diversity}

\subsection{Modeling replica diversity}
\label{sec:entropy}

We propose using entropy to measure replica diversity.  Let
$\D=\{d_1, \ldots, d_k\}$ be the complete space of replica
confirmation that can be remotely attested, where each element in $\D$
is a unique replica confirmation, s.t.  $d_i\neq d_j$ for all
$i,j\in [k]$, $[k]=\{1,2,...,k\}$ and $i\neq j$.

Let $\p = (p_1, \ldots, p_k)$ be a probability distribution of $\D$ on
$k$ replica configurations, i.e., $(d_1, \ldots, d_k)$. For all
$i \in [k]$, $p_i$ represents the ratio of replicas having
configuration $d_i$. If no replica has configuration $d_i$, then
$d_i=0$ and we define $\log \frac{1}{0}=0$. The Shannon entropy
$H(\p)$ of $\p$ is

\[
H(\p)
=
- \sum_{i \in [k]} p_i \log p_i
=
\sum_{i \in [k]} p_i \log \frac{1}{p_i}.
\label{entropy}
\]

For a given $\p$ of $\D$, the maximal value of $H(\p)$ represents the
ideal worst-case resilience, as maximising $H(\p)$ requires:
\begin{itemize}
\item \emph{Uniform distribution of replicas}. For the same number of
  available configuration (i.e., the number of nonzero elements in the
  probability distribution $\p$ is identical), $H(\p)$ is at its
  maximum when $p_i=p_j$, for all $i,j\in [k]$ and nonzero $p_i$ and
  $p_j$.
\item \emph{Greater diversity} in available replica configurations,
  i.e., more nonzero elements in $\p$.
\end{itemize}

The above two conditions to maximise entropy are orthogonal, as the
entropy of a probability distribution with more available replica
configurations is not necessarily larger. We define it formally below.

\begin{definition}
  ($\kappa$-optimal fault independence). For all $\kappa \leq k$, a
  replica configuration distribution $\p = (p_1, \ldots, p_k)$
  achieves $\kappa$-optimal fault independence iff the following
  holds:
  \begin{itemize}
  \item $|\p'|=\kappa$, where $\p'= \{\forall p_i\in \p: p_i\neq 0\}$;
  \item $\forall p_i,p_j\in \p'$, $\ p_i=p_j$.
  \end{itemize}
\end{definition}

\subsection{The impact of configuration abundance}
The definition of $\kappa$-optimal fault independence maximises the
entropy of distributions where $|\p'|=\kappa$. However, theoretically
they are not equivalent, which we will explain by using
\emph{configuration abundance}. In ecology, abundance has been used to
measure the number of individuals found per sample. In this work, we
use \emph{configuration abundance} to define the number of individuals
per replica configuration, and \emph{relative configuration abundance}
to represent the associated percent composition. The former is useful
for traditional BFT protocols, where the number of replica
matters. The latter is particularly useful for Bitcoin-like protocols,
where the relative configuration abundance represents mining power
distribution.

\begin{prop}\label{prop1}
  For $\kappa$-optimal fault independence system, increasing
  configuration abundance decreases entropy, unless the relative
  configuration abundance remains identical.
\end{prop}

\begin{prop}\label{prop2}
  Assuming each replica has a unique configuration, having more
  replicas does not provide more resilience, unless the relative
  configuration abundances are identical.
\end{prop}

\begin{figure}[!t]
  \centering
  \includegraphics[width=.9\columnwidth]{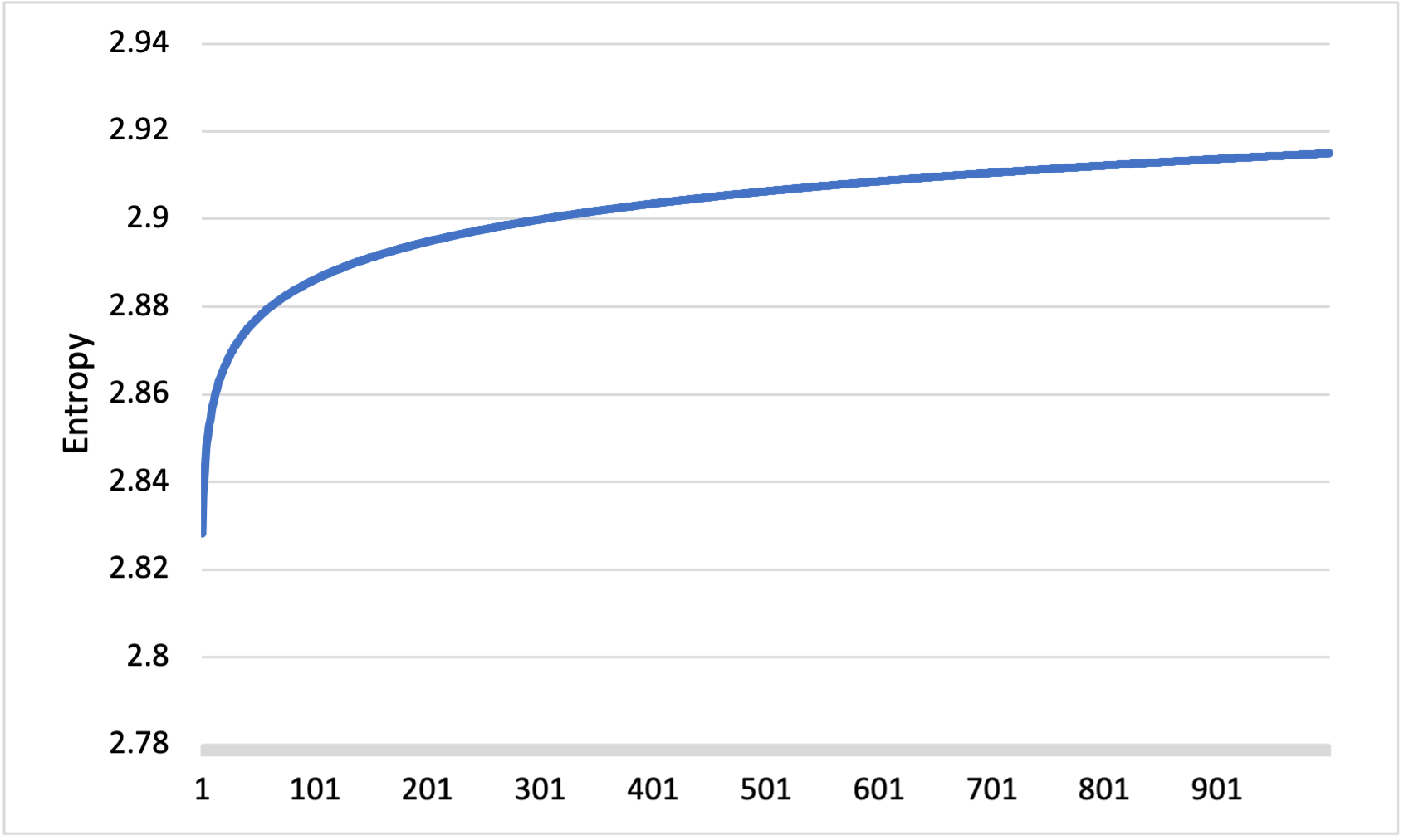}
  \caption{Best case entropy of Bitcoin replica diversity, assuming
    each miner has a unique configuration. X-axis indicates the number
    of miners where 0.87\% mining power has been uniformly distributed
    to. For example, when x=101, it means that there are 118 miners in
    the system, where the mining power (99.13\%) of the top 17 miners
    are distributed as in Example~\ref{example1} and the rest 0.87\%
    mining power are uniformly distributed among other 101 miners.}
  \label{fig:fig1}
\end{figure}

While Prop.~\ref{prop1} is intuitive, we use the following example to
illustrate Prop.~\ref{prop2}.
\begin{example}\label{example1}
  We analyse the best case entropy of replica diversity in Bitcoin
  (Figure~\ref{fig:fig1}). As of 02 February 2023, 17 mining pools in
  Bitcoin possess 99.13\% mining power, where the distribution is
  (34.239\%, 19.981\%, 12.997\%, 11.348\%, 8.826\%, 2.619\%, 2.037\%,
  1.649\%, 1.358\%, 1.261\%, 0.78\%, 0.68\%, 0.68\%, 0.39\%, 0.10\%,
  0.10\%,
  0.10\%)\footnote{https://www.blockchain.com/explorer/charts/pools
    (02 February 2023.)}. We take this as an example to quantify
  Bitcoin replica diversity. We consider the best possible
  configuration diversity of all replicas in Bitcoin, that it, we
  assume that each of the mining pools has a unique configuration
  where a fault in one pool does not affect other pools. In addition,
  since the distribution of the rest 0.87\% mining power is unknown,
  we consider the rest 0.87\% mining power is uniformly distributed to
  a number of replicas ranging from 1 to 1000. Figure~\ref{fig:fig1}
  presents the entropy distribution. This shows that, even with a
  large number of miners in Bitcoin, as there is an oligopoly in
  Bitcoin, the entropy is less than 3. However, when considering BFT
  protocols with 8 replicas, the entropy is already higher (entropy is
  3) also assuming a unique configuration per replica. This reveals
  that, at least from this simplified model, Bitcoin with a large
  number of miners does not provide better fault tolerance than BFT
  protocols with a small number of replicas, when oligopoly exists
  (represented by the relative configuration abundance).
\end{example}

Traditional BFT-SMR systems assume that each replica has a unique
configuration to guarantee optimal fault independence, i.e., the
configuration abundance is 1 for all configurations.  This is
sufficient but not necessary to achieve $\kappa$-optimal fault
independence, i.e., it satisfies our above $\kappa$-optimal fault
independence, but at the same time it also requires \emph{only one
  replica per unique configuration}. While this provides ideal fault
independence, i.e., the same fault does not allow an attacker to
control more than one replica, it only applies to the adversary model
where cyber-attackers controlling replicas via exploitable faults.

\begin{prop}\label{prop3}
  Higher configuration abundance improves the resilience of
  permissionless blockchains.
\end{prop}

This proposition is discussed below and motivates us to define optimal
resilience as follows:

\begin{definition}
  ($(\kappa, \omega)$-optimal resilience). A system is
  $(\kappa, \omega)$-optimal resilience if it is $\kappa$-optimal
  fault independence with configuration abundance of $\omega$.
\end{definition}

When Byzantine faults are only introduced by vulnerabilities (which is
more likely to be the case for permissioned systems), higher
configuration abundance doesn't help to improve resilience. However,
configuration abundance provides extra resilience in the case where
cyberattacks are not the only reason for having a Byzantine
replica. For example, a replica may choose to be malicious for gaining
financial benefits, which is a widely accepted adversary model and
concern in permissionless blockchains. In this case, a higher
configuration abundance help improve resilience as the malicious
operator cannot control other replicas of the same configuration. This
introduces a trade-off between performance and reliability, as a
higher configuration abundance always introduces more network
overhead, as the number of messages to communicate is also increasing
proportionally.

\section{Conclusion}
\label{sec:conclusion}
This paper investigated fault independence in permissionless
blockchains. We do not expect every replica to equip with a trusted
hardware for configuration attestation. However, having two types of
replicas (potentially with different voting right/weight), one
supporting configuration attestation and one does not, will help to
improve blockchain resilience. We leave further details and
discussions as future work.

\bibliographystyle{IEEEtran}
\bibliography{ref}

\end{document}